\documentclass{ifacconf}

\usepackage{graphicx}      
\usepackage[round,authoryear,sort]{natbib}        
\usepackage{graphics} 
\usepackage{epsfig} 
\usepackage{amsmath} 
\usepackage{amssymb}  
\usepackage{algorithm}
\usepackage{algpseudocode}
\usepackage{mathtools}

\usepackage{arydshln}

\newcommand{\set}[1]{\left\{#1\right\}}
\newcommand{\bs}{\boldsymbol}

\newcommand{\norm}[1]{\left\Vert#1\right\Vert}

\begin{document}
\begin{frontmatter}

\title{Numerical Discretization Methods \\ for Linear Quadratic Control Problems \\ with Time Delays} 


\author[]{Zhanhao Zhang},
\author[]{Steen  H{\o}rsholt},
\author[]{John Bagterp J{\o}rgensen}

\address{Department of Applied Mathematics and Computer Science \\ Technical University of Denmark, DK-2800 Kgs. Lyngby, Denmark (e-mail: \{jbjo\}@dtu.dk)}


\begin{abstract}
This paper presents the numerical discretization methods of the continuous-time linear-quadratic optimal control problems (LQ-OCPs) with time delays. We describe the weight matrices of the LQ-OCPs as differential equations systems, allowing us to derive the discrete equivalent of the continuous-time LQ-OCPs. Three numerical methods are introduced for solving proposed differential equations systems: 1) the ordinary differential equation (ODE) method, 2) the matrix exponential method, and 3) the step-doubling method. We implement a continuous-time model predictive control (CT-MPC) on a simulated cement mill system, and the objective function of the CT-MPC is discretized using the proposed LQ discretization scheme. The closed-loop results indicate that the CT-MPC successfully stabilizes and controls the simulated cement mill system, ensuring the viability and effectiveness of LQ discretization.
\end{abstract}


\end{frontmatter}

\section{Introduction}
Time delays are common in many industrial processes, shaping control systems’ trajectories by not solely relying on the present state but also integrating their history. These delays can significantly influence the robustness and control performance~\citep{LEE1995TimeDelaySystems,Gu2003StabilityOT}.The linear-quadratic optimal control problems (LQ-OCPs) with time delays find extensive practical applications and needs tailored algorithms for the system identification \citep{Jorgensen:Jorgensen:PEM:ACC2007,Jorgensen:Jorgensen:PEMforMPC:ACC2007,Jorgensen:Jorgensen:PEMforMPC:ECC2007} as well as in the linear quadratic Gaussian (LQG) control and model predictive control (MPC) implementation~\citep{BAGTERPJORGENSEN2012187,John2013EfficientImplementationofRiccati}. Therefore, there is a need for discretization methods tailored for LQ-OCPs with time delays.

Although research on discretization methods for time-delay systems is extensive~\citep{franklin1990DigitalControl,kassas2006discretization,Hendricks2008LCD,otto2017time}, studies on discretization methods for continuous-time LQ-OCP, particularly incorporating with time delays, are limited. \cite{Hendricks2008LCD} introduced the discrete-time approximation for LQ-OCPs without time delays, employing the zero-order hold (ZOH) parameterization on system states and inputs. \cite{Karl1970IntroToStoContTheory},~\cite{astrom2011computer} and~\cite{franklin1990DigitalControl} provided analytical expressions for equivalent discrete weighting matrices by extending continuous-time cost functions, which can be solved using the matrix exponential method~\citep{Moler1978NineteenDubiousWays,ExponentialIntegrators2011Higham}. The discretization and solution methods for continuous-time linear-quadratic regulator (CLQR) problems are described by~\cite{pannocchia2010computing,Pannocchia2015}. They used the matrix exponential method to obtain the discrete equivalent and proposed a novel computational procedure for solving the optimal control problem.

 On the other hand, stochastic LQ-OCPs can be valuable in some scenarios, such as Conditional Value-at-Risk (CVaR) optimization problems~\citep{Rockafellar2001Conditionalvalueatrisk,CAPOLEI2015214}. 
\cite{Karl1970IntroToStoContTheory} and~\cite{astrom2011computer} introduced the analytic expressions of continuous-time stochastic LQ-OCPs' cost function and described its expectation. However, as far as we know, the existing literature has not addressed the case of time delays or stochastic systems. In this paper, we thus focus on discretization methods for deterministic and stochastic LQ-OCPs with time delays. The key problems that we address in this paper: 
 \begin{itemize}
     \item [1.] Formulation of differential equation systems for LQ discretization with time delays
     \item [2.] Numerical methods for solving resulting systems of differential equations 
 \end{itemize}

This paper is organized as follows. Section~\ref{sec:LinearQuadraticOptimalControlProblems} introduces the discretization of time-delay systems and deterministic and stochastic LQ-OCPs with time delays. The discrete weighting matrices are described as differential equation systems. Section~\ref{sec:NumericalMethods} describes three numerical methods for solving proposed differential equation systems. We test the proposed numerical methods by a numerical experiment in Section~\ref{sec:NumericalExperiments} and give the conclusions in Section~\ref{sec:Conclusions}.
\section{Linear-Quadratic Optimal Control Problems}
\label{sec:LinearQuadraticOptimalControlProblems}
This section describes deterministic and stochastic LQ-OCPs with time delays. 
\subsection{Certainty equivalent LQ for deterministic time-delay systems}
Consider a SISO, time-delay, linear state space model
\begin{subequations}
\label{eq:Deterministic-ContinuousTimeDelayStateSpace}%
\begin{align} 
    \dot x(t) &= A_c x(t) + B_c u(t - \tau), \\ 
    z(t) &= C_c x(t) + D_c u(t - \tau),
\end{align} 
\end{subequations} 
where $x \in \mathbb{R}^{n_x \times 1}$ is the state, $u \in \mathbb{R}^{n_u \times 1} $ is the input and $z \in \mathbb{R}^{n_z \times 1}$ is the output. $\tau \in \mathbb{R}^{+}_0$ is the input time delay.

Assuming piece-wise constant input $u(t)=u_k$ for $t_k \leq t < t_{k+1}$ and $T_s$ is the sampling time. Note that $m \in \mathbb{Z}^+_0$ and $0 \leq v < 1$ are the integer and fractional time delay constants for $l=\tau/T_s=m-v$. By taking integral on both side, we obtain the solution of~\eqref{eq:Deterministic-ContinuousTimeDelayStateSpace} 
\begin{subequations}
    \label{eq:solutionOfSISODetStateSpace}
    \begin{align}
        x(t) & = A(t) x_k + B_o(t) \Tilde{u}_k,
        \\ 
        z(t) &= C_c x(t) + D_c u_{k-m} = C_c x(t) + D_o \tilde u_k,
    \end{align}
\end{subequations}
where $\tilde u_k=[u_{o,k};u_k]$ is the augmented input vector and $u_{o,k}=[u_{k-m};\cdots;u_{k-1}]$ is the historical input vector. 

The corresponding system matrices are
\begin{subequations}
\label{eq:SystemMatrices_solutionOfSISODetStateSpace}
\begin{align}
    & A(t) = e^{A_c t}, && B_1(t) = \int_0^t A(s) ds B_{1c}, \label{eq:AtAndB1t} 
    \\
    & A_v(t) = e^{v A_c t}, && B_2(t) =  v \int_0^t A_v(s) ds (B_{2c} - B_{1c}), \label{eq:AvtAndB2t} 
    \\
    & B_{1c}=B_c e^1_{m+1}, && B_{2c}=B_c e^2_{m+1}, 
    \\
    & D_o = D_c e^{1}_{m+1}, && B_o(t) = B_1(t) + B_2(t),
\end{align}    
\end{subequations}
where $e^p_{m+1} = [0 \cdots I \cdots 0]$ for $p=1,2,\ldots,m+1$ is an unit vector for selecting $u_{k-(m+1)+p}$ from $\tilde u_k$ such that $u_{k-{m+1}+p}=e^p_{m+1} \tilde u_k$. 


Set $t=T_s$, and we obtain
\begin{subequations}
\begin{align}
    \overbrace{ \left [
                    \begin{array}{c}
                        x_{k+1} \\ \hdashline[2pt/2pt]
                        u_{o,k+1}
                    \end{array}  \right]  }^{=\Tilde{x}_{k+1}} 
    &= \overbrace{ 
         \left[
            \begin{array}{c;{2pt/2pt}c}
                A & B_{o,1} \\ \hdashline[2pt/2pt]
                0 & I_A
            \end{array}
         \right] }^{=\Tilde{A}}
        \overbrace{ \left [
                    \begin{array}{c}
                        x_k \\ \hdashline[2pt/2pt]
                        u_{o,k}
                    \end{array}  \right]  }^{=\Tilde{x}_{k}}
        + 
        \overbrace{
             \left[
                \begin{array}{c}
                    B_{o,2} \\ \hdashline[2pt/2pt]
                    I_B
                \end{array}
             \right] }^{=\Tilde{B}} u_k
        , \\ 
    z_k &= \overbrace{
            \left[
                \begin{array}{c;{2pt/2pt}c}
                    C_c & D_{o,1}
                \end{array}
             \right]
            }^{=\Tilde{C}} \Tilde{x}_k + 
            \overbrace{D_{o,2}}^{=\Tilde{D}} u_k,
\end{align} 
\label{eq:Deterministic-DiscreteTimeDelayStateSpaceSISO}%
\end{subequations}
where the system matrices are
\begin{subequations} 
\label{eq:SystemMatricesOfDelayFreePresentationSISO} %
\begin{alignat}{3} 
    & B_{o,1} = B_o(:, 1:end-n_u), \, && 
    B_{o,2} = B_o(:, m n_u:end), 
    \\
    & D_{o,1} = D_o(:, 1:end-n_u), && 
    D_{o,2} = D_o(:, m n_u:end),
    \\
    & I_A = 
    \begin{bmatrix}
        0 & I & \ldots & 0 \\
        \vdots & \vdots & \ddots & \vdots \\
        0 & 0 & \ldots & I \\
        0 & 0 & \ldots & 0 
    \end{bmatrix},
    && I_B = 
    \begin{bmatrix}
        0 \\ \vdots \\ 0 \\  I 
    \end{bmatrix}. \label{eq:IAandIB}
\end{alignat} 
\end{subequations}
%
To present time delays in one single MIMO state space model, we consider the following $n_z \times n_u$ SISO systems
\begin{subequations}
\label{eq:Deterministic-ContinuousTimeDelayStateSpaceMIMO}
    \begin{align} 
        \dot x_{ij}(t) &= A_{c,ij} x_{ij}(t) + B_{c,ij} u_j(t - \tau_{ij}), \\ 
        z_{ij}(t) &= C_{c,ij} x_{ij}(t) + D_{c,ij} u_j(t - \tau_{ij}), 
    \end{align}
with 
\begin{align}
    & \quad u=[u_1;u_2;\ldots,u_{n_u}],  &&x=[x_{11};x_{21};\ldots;x_{n_zn_u}],
    \\
    & \quad z=[z_{1};z_{2};\ldots;z_{n_z}], && z_i = \sum_{j=1}^{n_u} z_{ij},
\end{align}
\end{subequations}
where $A_{c,ij}$, $B_{c,ij}$, $C_{c,ij}$, $D_{c,ij}$ and $\tau_{ij}$ for $i=1,2,\ldots,n_z$ and $j = 1,2,\ldots, n_u$ are parameters of the $[i, j]$ SISO system. It describes the dynamics from the j$^{th}$ input to i$^{th}$ output. The corresponding time delay constants are denoted as $l_{ij} = m_{ij} - v_{ij} = \tau_{ij}/T_s$.

The historical input vector is $u_{o,k}=[u_{k-\bar m}; \ldots; u_{k-1}]$ with $\bar m = \max{\set{m_{ij}}}$. For the augmented input vector $\tilde u$, we have the following expression in the MIMO case
\begin{equation}
    u_{j,k-(\bar m+1)+p} = e_j u_{k-(\bar m+1)+p} = e_j E_{\bar m+1}^p \tilde u_k,
\end{equation}
where $E_{\bar m+1}^p$ is an unit matrix that select $u_{k-(\bar m +1)+p}$ from $\tilde u_k$ and $e_j=[0 \cdots 1 \cdots 0]$ for $j=1,2,\ldots,n_u$ is an unit vector for selecting $j^{th}$ input from $u_{k-(\bar m+1)+p}$. 

Stacking all SISO systems' solutions, we can obtain the solution of the MIMO time-delay model that has the same expressions as the SISO case~\eqref{eq:solutionOfSISODetStateSpace}. The matrices $A$, $A_v$, $B_1$, $B_2$ have the same expressions introduced in~\eqref{eq:AtAndB1t} and~\eqref{eq:AvtAndB2t}, and their coefficients become
%
\begin{subequations}
    \begin{align}
    & A_c = \text{diag}(A_{c,11},A_{c,21},\ldots,A_{c,n_zn_u}), 
    \\
    & B_{1c} = [B_{1c,11};B_{1c,21};\ldots;B_{1c,n_zn_u}], 
    \\
    & B_{2c} =  [B_{2c,11};B_{2c,21};\ldots;B_{2c,n_zn_u}],
    \\
    &  V=\text{diag}(V_{11},V_{21},\ldots,V_{n_zn_u}), 
    \end{align}
\end{subequations}
where $V_{ij} = I v_{ij}$, $B_{1c,ij} = B_{c,ij} e_j E_{\bar m+1}^{m_{ij}}$ and $B_{2c,ij}=B_{c,ij} e_j E_{\bar m+1}^{m_{ij}+1}$.

The matrices $C_c$ and $D_o$ output function matrices become 
\begin{subequations}
    \begin{align}
         & C_c = [\bar C_1, \bar C_2, \cdots, \bar C_{n_u}], 
        \\
        & D_o = [\bar D_{1}; \bar D_2; \cdots; \bar D_{n_z}],
        \\
        & \bar{C}_j=\text{diag}(C_{c,1j}, C_{c,2j}, \ldots,C_{c,n_zj}),
        \\
        & \bar{D}_{c,i} = \sum_{j=1}^{n_u} D_{c,ij} e_j E_{\bar m+1}^{m_{ij}}, \quad \text{for} \: i=1,2,\ldots,n_z.
    \end{align}
\end{subequations}


Consequently, set $t=T_s$, we can obtain the discrete-time system of~\eqref{eq:Deterministic-ContinuousTimeDelayStateSpaceMIMO} with same expressions introduced in the SISO case~\eqref{eq:Deterministic-DiscreteTimeDelayStateSpaceSISO} with $B_{o,1} = B_o(:, 1:end-n_u)$, $B_{o,2} = B_o(:, \bar{m} n_u:end)$, $D_{o,1} = D_o(:, 1:end-n_u)$ and $D_{o,2} = D_o(:, \bar{m} n_u:end)$.
\begin{prop}
\label{prop:DiscretizationofMIMOTimeDelaySystem}
The system of differential equations
\begin{subequations}
\label{eq:MIMOTimeDelaySystem:MatrixODEsystem}
\begin{alignat}{3}
    \dot A(t) &= A_c A(t), \qquad && A(0) = I, \\
    \dot A_v(t) &= V A_c A_v(t), \qquad && A_v(0) = I, \\ 
    \dot B_1(t) &= A(t) B_{1c}, \qquad && B_1(0) = 0, \\
    \dot B_2(t) &= A_v(t) \bar{B}_{2c}, \qquad && B_2(0) = 0,
\end{alignat}
\end{subequations}
where 
\begin{equation}
    \bar{B}_{2c} = V(B_{2c} - B_{1c}),
\end{equation}
may be used to compute ($A = A(T_s)$, $B_o=B_1(T_s)+B_2(T_s)$) for the discretization of MIMO time-delay systems.
\end{prop}
\subsection{Deterministic linear-quadratic optimal control problem}
Consider the following deterministic LQ-OCP
\begin{subequations}
\label{eq:Deterministic:ContinuousTime:LinearQuadraticOCP}
\begin{alignat}{5}
& \min_{x,u,z,\tilde z} &&\phi = \int_{t_0}^{t_0+T} l_c(\tilde z(t)) dt \\
& s.t. && x(t_0) = \hat x_0, \\
& && u(t) = u_k, \: t_k \leq t < t_{k+1}, \: k \in \mathcal{N}, \\
& && \dot x(t) = A_c x(t) + B_c u(t-\tau), t_0 \leq t < t_0+T, \\
& && z(t) = C_c x(t) + D_c u(t-\tau), t_0 \leq t < t_0+T, \\
& && \bar z(t) = \bar z_k, \: t_k \leq t < t_{k+1}, \: k \in \mathcal{N}, \\
& && \tilde z(t) = z(t) - \bar z(t), \: t_0 \leq t < t_0+T,
\end{alignat}
\end{subequations}
with the stage cost function
\begin{equation}
\begin{split}
    l_c(\tilde z(t)) &= \frac{1}{2} \norm{ W_z \tilde z(t)  }_2^2
    = \frac{1}{2} \tilde z(t) ' Q_{c} \tilde z(t), 
\end{split}
\label{eq:det-stagecost}
\end{equation}
where $T = N T_s$ and $N \in \mathbb{Z}^+$ for $\mathcal{N} = 0, 1, \ldots, N-1$ is the control interval. $Q_c=W_z' W_z$ is a semi-positive definite weight matrix. 

We assume piece-wise constant inputs $u(t) = u_k$ and target variables $\bar z(t) = \bar{z}_k$ for $t_k \leq t < t_{k+1}$. Replacing $z(t)$ with the expressions introduced in~\eqref{eq:solutionOfSISODetStateSpace}, the discrete-time equivalent of~\eqref{eq:Deterministic:ContinuousTime:LinearQuadraticOCP} can be defined as
\begin{subequations}
\label{eq:Deterministic:DiscreteTime:LinearQuadraticOCP}
\begin{alignat}{5}
& \min_{x,u} \quad &&\phi = \sum_{k\in \mathcal{N}} l_k(x_k,u_k)  \\
& s.t. && x_0 = \hat x_0, \\
& && x_{k+1} = A x_k + B u_k, \quad && k \in \mathcal{N},
\end{alignat}    
\end{subequations}
where the states $x$ and system matrices $A$, $B$, $C$ and $D$ are in the augmented form and correspond  to $\tilde x, \tilde A, \tilde B, \tilde C, \tilde D$ described in~\eqref{eq:Deterministic-DiscreteTimeDelayStateSpaceSISO}. 

The stage cost function $l_k(x_k, u_k)$ is 
\begin{equation}
    l_k(x_k,u_k) = \frac{1}{2} \begin{bmatrix} x_k \\ u_k \end{bmatrix}' Q \begin{bmatrix} x_k \\ u_k \end{bmatrix} + q_k' \begin{bmatrix} x_k \\ u_k \end{bmatrix} + \rho_k, \quad k \in \mathcal{N},
\label{eq:deterministic-stageCost}
\end{equation}
and its affine term's coefficients and the constant term are 
\begin{equation}
    q_k =  M \bar{z}_k, \:  \rho_k =  \int_{t_k}^{t_{k+1}} l_c(\bar{z}_k) dt = l_c(\bar z_k) T_s, \: k \in \mathcal{N}.
\label{eq:qkandrho_k}
\end{equation}
%
\begin{prop} The system of differential equations
\label{prop:DiscretizationoftheDterministicLQOCP}
\begin{subequations}
\label{eq:DeterministicLQ:MatrixODEsystem}
\begin{alignat}{3}
    \dot A(t) &= A_c A(t), \qquad && A(0) = I, \\
    \dot A_v(t) &= VA_c A_n(t), \qquad && A_v(0) = I, \\
    \dot B_1(t) &=  A(t)  B_{1c}, \qquad && B_1(0) = 0, \\
    \dot B_2(t) &= A_v(t) \bar B_{2c}, \qquad && B_2(0) = 0, \\
    \dot Q(t) &= \Gamma(t)'  Q_{c} \Gamma(t), \qquad && Q(0) = 0,  \\
    \dot M(t) &= -\Gamma(t)' M_{c}, \qquad && M(0) = 0,
\end{alignat}
\end{subequations}
where
\begin{equation}
    \bar{B}_{2c} = V(B_{2c} - B_{1c}), \: \:
    \Gamma(t) = \begin{bmatrix} C_c & D_o \end{bmatrix} \begin{bmatrix} A(t) & B_o(t) \\ 0 & I \end{bmatrix}, 
\end{equation}
may be used to compute ($A = A(T_s)$, $B_o=B_1(T_s)+B_2(T_s)$, $Q=Q(T_s)$, $M=M(T_s)$) for the discretization of deterministic LQ-OCPs with time delays.
\end{prop}
\subsection{Certainty equivalent LQ control for a stochastic time-delay system}
Consider the following linear, stochastic, time-delay system
\begin{subequations}
\label{eq:linearSDE}
\begin{alignat}{3}
    d {\bs x}(t) &= \left( A_c {\bs x}(t) + B_c u(t-\tau) \right) dt + G_c d{\bs \omega}(t), \\
    {\bs z}(t) &= C_c {\bs x}(t) + D_c  u(t-\tau).
\end{alignat}
\end{subequations}
and the initial state $\bs x_0 \sim N(\hat x_0,P_0)$ and stochastic variable $d{\bs \omega} (t) \sim N_{iid}(0, I dt)$.

Based on expressions obtained in the deterministic case~\eqref{eq:solutionOfSISODetStateSpace} and~\eqref{eq:Deterministic-DiscreteTimeDelayStateSpaceSISO}, we can define the discrete-time system of~\eqref{eq:linearSDE} as
\begin{subequations}
\label{eq:Stochastic-DiscreteTimeDelayStateSpaceMIMO}%
\begin{alignat}{3}
    \Tilde{\bs x}_{k+1} &= \Tilde{A} \Tilde{\bs x}_k + \Tilde{B} u_k + \Tilde{\bs w}_k, \\
    {\bs z}_k &= \Tilde{C} \Tilde{\bs x}_k +  \Tilde{D} u_k,
\end{alignat} 
and $\tilde{\bs w}_k=[\bs w_k; 0]$ is expressed in terms of It$\hat{o}$
\begin{equation}
    \bs w_k = \int_{t_k}^{t_{k+1}} A(t) G_c d \bs \omega(t), 
    \quad 
    \bs w_k \sim N_{iid} \left( 0, R_{ww} \right),
\end{equation}
\end{subequations}
where $\Tilde{A}$, $\Tilde{B}$, $\Tilde{C}$ and $\Tilde{D}$ are identical to the deterministic case~\eqref{eq:Deterministic-DiscreteTimeDelayStateSpaceSISO} and $R_{ww} = \text{Cov}(\bs w_k)$ is the covariance matrix. 
\begin{prop}
\label{prop:DiscretizationoftheLinearSDE}
The system of differential equations
\begin{subequations}
\begin{alignat}{3}
    \dot A(t) &= A_c A(t), \qquad && A(0) = I, \\
    \dot A_v(t) &= V A_c A_v(t), \qquad && A_v(0) = I, \\ 
    \dot B_1(t) &= A(t) B_{1c}, \qquad && B_1(0) = 0, \\
    \dot B_2(t) &= A_v(t) \bar{B}_{2c}, \qquad && B_2(0) = 0, \\
    \dot R_{ww} &= \Phi(t) \Phi(t)', \qquad && R_{ww}(0) = 0,
\end{alignat}
\end{subequations}
where 
\begin{equation}
\bar{B}_{2c} = V(B_{2c} - B_{1c}), \quad \Phi(t) = A(t) G_c,
\end{equation}
may be used to compute ($A=A(T_s)$, $B_o=B_1(T_s)+B_2(T_s)$, $R_{ww}=R_{ww}(T_s)$) for the discretization of stochastic time-delay models.
\end{prop}
\subsection{Stochastic linear-quadratic optimal control problem}
Consider the stochastic LQ-OCP governed by~\eqref{eq:linearSDE}
\begin{subequations}
\label{eq:Cont-Stc-LQ-OCP}
\begin{alignat}{5}
& \min_{{\bs x},u,{\bs z},{\tilde{\bs z}}} \: \psi = E \set{ \phi = \int_{t_0}^{t_0+T} l_c({\tilde{\bs z}}(t))dt }  \MoveEqLeft[1] \\
&  s.t. \quad  {\bs x}(t_0) \sim N(\hat x_0, P_0), \\
&  \qquad \: d{\bs \omega}(t) \sim N_{iid}(0,I dt), \\
&  \qquad \: u(t) = u_k, \quad t_k \leq t < t_{k+1}, \, k \in \mathcal{N}, \\
& \qquad \: d{\bs x}  (t) = (A_c {\bs x}(t) + B_c u(t-\tau))dt + G_c d{\bs \omega}(t), \\
& \qquad \: {\bs z}(t) = C_c {\bs x}(t) + D_c u(t-\tau), \\
& \qquad \: {\bar z}(t) = \bar z_k, \quad t_k \leq t < t_{k+1}, \, k \in \mathcal{N}, \\
& \qquad \: {\tilde {\bs z}}(t) = {\bs z}(t) - \bar{z}(t).
\end{alignat}
\end{subequations}
The corresponding discrete-time stochastic LQ-OCP is 
\begin{subequations}
\label{eq:Dist-Stc-LQ-OCP}
\begin{alignat}{5}
& \min_{ x,u} \quad  && \psi = E\set{\phi = \sum_{k \in \mathcal{N}} l_k({\bs x}_k,u_k) + l_{s,k} (\bs{x}_k,u_k)}  \\ 
& s.t. && {\bs x}_0 \sim N(\hat x_0, P_0), \\
& && {\bs w_k} \sim N_{iid}(0,R_{ww}), \qquad \qquad k \in \mathcal{N}, \\
& && {\bs x}_{k+1} = A {\bs x}_k + B u_k + {\bs w}_k, \quad \: k \in \mathcal{N}, 
\end{alignat}
\end{subequations}
where the variables $\bs x$, $\bs w$ and system matrices $A$, $B$, $C$, $D$ correspond to $\tilde{\bs{x}}, \tilde{\bs{w}},\tilde A, \tilde B, \tilde C, \tilde D$ described in~\eqref{eq:Stochastic-DiscreteTimeDelayStateSpaceMIMO}. 

The stage cost function $l_k(\bs{x}_k, u_k)$ is
\begin{equation}
l_k(\bs{x}_k,u_k) = \frac{1}{2} \begin{bmatrix} \bs{x}_k \\ u_k \end{bmatrix}' Q \begin{bmatrix} \bs{x}_k \\ u_k \end{bmatrix} + q_k' \begin{bmatrix} \bs{x}_k \\ u_k \end{bmatrix} + \rho_k,
\label{eq:deterministicStageCostofStochasticCosts}
\end{equation}
and $l_{s,k}(\bs{x}_k, u_k)$ is 
\begin{equation}
    l_{s,k}(\bs{x}_k, u_k) = \int_{t_k}^{t_{k+1}} \frac{1}{2} \bs{w}(t)' Q_{c,ww} \bs{w}(t) + \bs{q}_{s,k}' \bs{w}(t) dt,
\end{equation}
where $Q$, $q_k$, and $\rho_k$ are identical to the deterministic case described in~\eqref{eq:qkandrho_k} and~\eqref{eq:DeterministicLQ:MatrixODEsystem}. The state $\bs w(t)$ and system matrices $Q_{c,ww}$ and $\bs q_{s,k}$ of $l_{s,k}(\bs{x}_k,u_k)$ are
\begin{subequations}
    \begin{align}
        & \bs{w}(t) = \int_{0}^{t} A(s) G_c d\bs{\omega}(s),  \quad 
        && Q_{c,ww} = C_c' Q_{c} C_c,  
        \\
        & \tilde{\bs{z}}_k = \Gamma(t) \begin{bmatrix} \bs{x}_k \\ u_k \end{bmatrix} - \bar{z}_k, 
        && \bs{q}_{s,k} = C_c' Q_{c} \Tilde{\bs{z}}_k.
    \end{align}
\end{subequations}
Based on the previous work by~\cite{Karl1970IntroToStoContTheory}, we can rewrite~\eqref{eq:Dist-Stc-LQ-OCP} as
\begin{subequations} 
\label{eq:analyticMeanOfStochasticCosts01}
\begin{alignat}{5}
& \min_{ x,u } \: && \psi = \sum_{k \in \mathcal{N}} l_k(x_k,u_k) + \bs \rho_{s,k}\\
& s.t. && x_0 = \hat x_0, \\
& &&  x_{k+1} = A x_k + B u_k, \quad k \in \mathcal{N}, 
\end{alignat}
\end{subequations}
where $\bs \rho_{s,k}$ is 
\begin{subequations} \label{eq:analyticMeanOfStochasticCosts02}
    \begin{equation}
        \bs \rho_{s,k} = \frac{1}{2} \left[ \text{tr}\left( Q \bar{P}_k \right) + \int_{t_k}^{t_{k+1}} \text{tr} \left(Q_{c,ww} P_{w} \right) dt \right],
    \end{equation}
and $ $
    \begin{equation}
    \begin{bmatrix}
            \bs{x}_k \\ u_k
        \end{bmatrix} \sim N(m_k, \bar{P}_k),
    \: 
    m_k = \begin{bmatrix}
        x_k \\ u_k
    \end{bmatrix},
    \: 
    \bar{P}_k = \begin{bmatrix} P_k & 0 \\ 0 & 0 \end{bmatrix},
\end{equation}
\begin{equation}
    P_{k+1} = A P_k A' + R_{ww}, \qquad P_w = \text{Cov} \left(\bs{w}(t)\right).
\end{equation}
\end{subequations}
\begin{prop}
\label{prop:DiscretizationoftheStochasticLQ-OCP}
The system of differential equations
\begin{subequations}
\label{eq:StochasticLQG:MatrixODEsystem}
\begin{alignat}{3}
    \dot A(t) &= A_c A(t), \qquad && A(0) = I, \\
    \dot A_v(t) &= V A_c A_v(t), \qquad && A_v(0) = I, \\ 
    \dot B_1(t) &= A(t) B_{1c}, \qquad && B_1(0) = 0, \\
    \dot B_2(t) &= A_v(t) \bar{B}_{2c}, \qquad && B_2(0) = 0, \\
    \dot Q(t) &= \Gamma(t)' Q_{c} \Gamma(t), \qquad && Q(0) = 0,  \\
    \dot M(t) &= -\Gamma(t)' Q_{c}, \qquad && M(0) = 0, \\
    \dot R_{ww}(t) &= \Phi(t) \Phi(t)', \qquad && R_{ww}(0) = 0, 
\end{alignat}
\end{subequations}
where 
\begin{subequations}
\begin{equation}
   \bar{B}_{2c} = V(B_{2c} - B_{1c}), \qquad \Phi(t) = A(t)G_c,
\end{equation}
\begin{equation}
    \Gamma(t) = \begin{bmatrix} C_c & D_o \end{bmatrix} \begin{bmatrix} A(t) & B_o(t) \\ 0 & I \end{bmatrix}, \quad 
\end{equation}
\end{subequations}
may be used to compute ($A=A(T_s)$, $B_o=B_1(T_s)+B_2(T_s)$, $Q=Q(T_s)$, $M=M(T_s)$, $R_{ww}=R_{ww}(T_s)$) for the discretization of stochastic LQ-OCPs with time delays.
\end{prop}
\section{Numerical Discretization Methods}
\label{sec:NumericalMethods}
In this section, we introduce numerical methods for solving proposed differential equations systems.
\subsection{Ordinary differential equation methods}
\begin{algorithm}[tb]
\caption{ODE method for LQ Discretization}
\label{alg:ODEmethod-LQDiscretization} %
\begin{flushleft}
    \textbf{Input:} $(A_c, B_c, G_c, C_c, D_c, Q_c, T_s, N)$ \\
    \textbf{Output:} $(A,B,Q,M,R_{ww})$ 
\end{flushleft}
\begin{algorithmic}
\State Set initial states \\
        ($k=0$, $A_k = I$, $A_{v,k}=I$, $B_{1,k} = 0$, $B_{2,k} = 0$, $Q_k = 0$, $M_k = 0$, $R_{ww, k} = 0$)
\State Compute the step size $h = \frac{T_s}{N}$
\State Use~\eqref{eq:ODEmethods-StageVariableCoefficients} to compute ($\Lambda_i$, $\Lambda_{v,i}$, $\Theta_{1,i}$, $\Theta_{2,i}$)
\State Use~\eqref{eq:ODEmethods-ConstantCoefficients} to compute ($\Lambda$, $\Lambda_{v}$, $\Theta_{1}$, $\Theta_{2}$)
\While{$k < N$}
    \State {Use~\eqref{eq:ODEmethods-numericalExpressions} to update ($A_{k}$, $B_{o,k}$, $Q_{k}$, $M_{k}$, $R_{ww, k}$)} 
    \State {Set $k \leftarrow k + 1$}
\EndWhile
\State Get system matrices
 $(A(T_s) = A_k, B(T_s)= B_k, Q(T_s) = Q_k, M(T_s) = M_k, R_{ww}(T_s) = R_{ww, k})$
\end{algorithmic} %
\end{algorithm}
Consider an s-stage fixed-time-step ODE method with $N \in \mathbb{Z}^+$ integration steps and the time step $h = \frac{T_s}{N}$. Define $a_{i,j}$ and $b_i$ for $i=1,2,\ldots,s$ and $j=1,2,\ldots,s$ are the Butcher tableau parameters of the ODE method, and we can compute ($A$,$B_{o}$,$Q$,$M$,$R_{ww}$) as 
\begin{subequations}
    \label{eq:ODEmethods-numericalExpressions}
    \begin{alignat}{3}
        & A_{k+1} = \Lambda A_k, \: 
        &&  B_{1, k+1} = B_{1,k} + \Theta_1 A_k \bar{B}_{1c},
        \\
        & A_{v,k+1} = \Lambda_v A_{v, k}, \: 
        &&
        B_{2,k+1} = B_{2,k} + \Theta_2 A_{v,k} \bar{B}_{2c},
        \\
        & B_{o,k} = B_{1,k}+B_{2,k}, \; \; &&
        \Gamma_{k+1} = \begin{bmatrix}
            A_{k+1} & B_{o,k+1} \\ 0 & I
        \end{bmatrix},
    \end{alignat}
    \begin{alignat}{3}
        &  M_{k+1} = M_k + h\sum_{i=1}^s b_i \Gamma_{k,i}' \bar{M}_c,
        \\
        &  Q_{k+1} = Q_k + h\sum_{i=1}^s b_{i} \Gamma_{k,i}' \bar Q_c \Gamma_{k,i},
        \\
        & R_{ww,k+1} = R_{ww,k} + h \sum_{i=1}^{s} b_i A_{k, i} \bar{R}_{ww,c} A_{k, i}'.
    \end{alignat}
\end{subequations}
 $\bar{B}_{1c}=h B_{1c}$, $\bar{B}_{2c}=hV \left( B_{2c} - B_{1c} \right)$, $\bar M_c$=$-\begin{bmatrix} C_c & D_o \end{bmatrix}' Q_c$, $\bar Q_c$=$-M_c \begin{bmatrix} C_c & D_o \end{bmatrix}$ and $\bar{R}_{ww,c}$=$G_c G_c'$ are constant. Note that $A_{k,i}$, $A_{v,k,i}$, $B_{1,k,i}$, $B_{2,k,i}$ and $\Gamma_{k,i}$ for $i$=$1,2,\ldots,s$ are stage variables of ($A$, $A_v$, $B_1$, $B_2$, $\Gamma$), we then have 
\begin{subequations}
    \label{eq:ODEmethods-StageVariableCoefficients}
    \begin{alignat}{3}
        & A_{k,i} = A_k + h\sum_{j=1}^{s} a_{i, j} \dot A_{k,j} = \Lambda_{i} A_k, \MoveEqLeft[1]
        \\
        & A_{v,k,i} = A_{v,k} + h\sum_{j=1}^{s} a_{i, j} \dot A_{v,k,j} = \Lambda_{v,i} A_{v,k}, \vspace{2.0cm}
    \end{alignat}
    \begin{equation}
        B_{1,k,i} = B_{1,k} + h\sum_{j=1}^{s} a_{i,j} \dot B_{1,k,j} = B_{1,k} + \Theta_{1,i} A_k \bar B_{1c},
    \end{equation}
    \begin{equation}
        B_{2,k,i} = B_{2,k} + h\sum_{j=1}^{s} a_{i,j} \dot B_{2,k,j} = B_{2,k} + \Theta_{2,i} A_{v,k} \bar B_{2c},
    \end{equation}
    \begin{equation}
        B_{o,k,i} = B_{1,k,i} + B_{2,k,i}, \quad 
        \Gamma_{k,i} = \begin{bmatrix}
        A_{k,i} & B_{o,k,i} \\ 0 & I 
    \end{bmatrix},
    \end{equation}
\end{subequations}
and we can compute coefficients $\Lambda$, $\Lambda_v$, $\Theta_1$, and $\Theta_2$ as
\begin{subequations}
    \label{eq:ODEmethods-ConstantCoefficients}
    \begin{alignat}{3}
        & \Lambda = I + h \sum_{i=1}^s b_i A_c \Lambda_i, \quad 
        && \Theta_1 = \sum_{i=1}^s b_i\Lambda_{i}, 
        \\ 
        & \Lambda_v = I + h \sum_{i=1}^s b_i V A_c \Lambda_{v,i},  \quad
        && \Theta_2 = \sum_{i=1}^s b_i\Lambda_{v,i}.
    \end{alignat}
\end{subequations}
where the stage variable coefficients $\Lambda_i$, $\Lambda_{v,i}$, $\Theta_{1,i}$ and $\Theta_{2,i}$ are functions of Butcher tableau's parameters $a$ and $b$. 

In particular, $\Gamma$ can be decomposed into the linear combination of $A$, $A_v$, $B_1$ and $B_2$, such that 
\begin{subequations}
\label{eq:Gammak}
\begin{equation}
    \begin{split}
        \Gamma(t) & = \begin{bmatrix}
            A(t) & B_o(t) \\ 0 & I 
        \end{bmatrix}
        \\
        & = \overbrace{\begin{bmatrix}
            A(t) & B_1(t) \\ 0 & I 
        \end{bmatrix}}^{H_1(t)} + \overbrace{\begin{bmatrix}
            A_v(t) & B_2(t) \\ 0 & I 
        \end{bmatrix}}^{H_2(t)} - \overbrace{\begin{bmatrix}
            A_v(t) & 0 \\ 0 & I 
        \end{bmatrix}}^{H_3(t)}
        \\
        & = E_1 H(t) E_2,
    \end{split}
\end{equation}
and 
\begin{align}
    & \qquad H_{1c} = \begin{bmatrix} A_c & B_{1c} \\ 0 & 0 \end{bmatrix}, \qquad 
    H_{2c} = \begin{bmatrix} \bar A_{c} & \bar B_{2c} \\ 0 & 0 \end{bmatrix}, 
    \\
    & H_{3c} = \begin{bmatrix} A_{v,c} & 0  \\ 0 & 0 \end{bmatrix}, \: \:  E_1 = [I, I, -I], \: \: 
    E_2 = [I; I; I],
\end{align}
\end{subequations}
where $H(t)=\text{diag}(H_1(t),H_2(t),H_3(t))=e^{H_c t}$ for $H_c=\text{diag}(H_{1c}, H_{2c}, H_{3c})$ and $H_k(t)=e^{H_{kc}t}$ for $k=1,2,3$.


Consequently, we compute $A(T_s)=A_N$, $B_o(T_s)=B_{o,N}$, $M(T_s)=M_{N}$, $Q(T_s) = Q_{N}$ and $R_{ww}(T_s) = R_{ww,N}$ with constant coefficients $\Lambda$, $\Lambda_v$, $\Theta_{1}$ and $\Theta_2$ using fixed-time-step ODE methods. Algorithm~\ref{alg:ODEmethod-LQDiscretization} describes the fixed-time-step ODE method for the LQ discretization with time delays.

\subsection{Matrix exponential method}
Based on formulas introduced by~\cite{vanLoan1978MatrixExponential} and~\cite{Moler1978NineteenDubiousWays,Moler2003NineteenDubiousWays25YearsLater}, we can discretize the LQ-OCP with time delays by solving the following matrix exponential problems
\begin{subequations}
\label{eq:matrixExponentialS}
    \begin{alignat}{5}
        \begin{bmatrix}
            \Phi_{1, 11} & \Phi_{1, 12}\\
            0 & \Phi_{1,22}
        \end{bmatrix} &= \text{exp} \left(\begin{bmatrix}
                            -H_c' & E_1' \bar Q_c E_1 \\
                            0   & H_c
                        \end{bmatrix}t \right), 
        \\
        \begin{bmatrix}
            I & \Phi_{2, 12}\\
            0 & \Phi_{2,22}
        \end{bmatrix} &= \text{exp} \left(\begin{bmatrix}
                            0 & I \\
                            0 & H_c'
                        \end{bmatrix}t \right), 
        \\
        \begin{bmatrix}
            \Phi_{3, 11} & \Phi_{3, 12}\\
                0 & \Phi_{3,22}
        \end{bmatrix} &= \text{exp} \left(\begin{bmatrix}
                            -A_c & \bar R_{ww,c} \\
                            0   & A_c'
                        \end{bmatrix} t \right), 
    \end{alignat}
\end{subequations}


The elements of matrix exponential problems~\eqref{eq:matrixExponentialS} are 
\begin{subequations}
    \begin{align}
        \begin{split}
            \Phi_{1,22} &= H(t) = \text{diag} \left( \begin{bmatrix}
           H_1(t) & H_2(t) & H_3(t)
            \end{bmatrix} \right),
        \end{split}\\
        \begin{split}
            \Phi_{1,12} & = H(-t)' \int_{0}^{t} H(\tau)' E_1' \bar{Q}_c E_1 H(\tau) d\tau,
        \end{split}\\
        \begin{split}
            \Phi_{2,12} &= \int_{0}^{t} H(\tau)' d\tau,
        \end{split}\\
        \begin{split}
            \Phi_{3,22} & = A(t),
        \end{split}\\
        \begin{split}
            \Phi_{3,12} &= A(-t) \int^{t}_{0} A(\tau) \bar{R}_{ww,c} A(\tau)' d \tau.
        \end{split}
    \end{align}
\end{subequations}
where $H_c$, $\bar Q_c$, $\bar M_c$, $\bar R_{ww,c}$ are introduced in~\eqref{eq:ODEmethods-numericalExpressions} and~\eqref{eq:Gammak}.

Consequently, set $t=T_s$, we can compute ($A$, $B_o$, $Q$, $M$, $R_{ww}$) as 
\begin{subequations}
    \begin{align}
        A(T_s) &= \Phi_{1,22}(1:n_x, 1:n_x), \\
        B_o(T_s) &= \Phi_{1,22}(1:n_x, n_x+1:\text{end}), \\
        \Gamma(T_s) &= E_1 \Phi_{1,22} E_2, \\
        Q(T_s) &= E_2' \Phi_{1,22}'\Phi_{1,12} E_2, \\
        M(T_s) &= E_2' \Phi_{2,12} \bar{M}_c, \\
        R_{ww}(T_s) &= \Phi_{3,22}' \Phi_{3,12}.
    \end{align}
\end{subequations}
\subsection{Step-doubling method}
Consider the matrix $H(t)=e^{H_ct}$, and we can express it in the form of the differential equation as  
\begin{equation}
    \dot{H}(t) = H_c H(t), \qquad H(0) = I_h,
\end{equation}
and its ODE expressions are 
\begin{subequations}
\label{eq:HkandGammak}
    \begin{alignat}{3}
        & H_{k+1} = \Omega H_k, \quad 
        &&
        \Gamma_{k} = \begin{bmatrix}
            A_k & B_{o,k} \\ 0 & I
        \end{bmatrix} = E_1 H_k E_2,
        \\
        & H_{k, i} = \Omega_i H_{k},
        && 
        \Gamma_{k, i} = \begin{bmatrix}
            A_{k,i} & B_{o,k,i} \\ 0 & I 
        \end{bmatrix} = E_1 H_{k,i} E_2,
    \end{alignat}
\end{subequations}
where $I_h$ is an identity matrix that has the same dimension as $H_c$. $H_{k,i}$ for $i=1,2,\ldots,s$ indicate the stage variables of $H(t)$ and their coefficients $\Omega_i$ are functions of Butcher tableau's parameters $a_{i,j}$ and $b_i$, such that $\Omega = I + h \sum_{i=1}^s b_i H_c \Omega_i$.

Consider the ODE expressions~\eqref{eq:ODEmethods-numericalExpressions} and~\eqref{eq:HkandGammak}, the matrices
\begin{subequations}
\label{eq:stepDoublingMatices}
    \begin{alignat}{3}
        & \Tilde{A}(N) = \bar{\Lambda}^N, && \Tilde{A}(1) = \bar{\Lambda}, 
        \\
        & \Tilde{B}_o(N) = \sum_{k=0}^{N-1} \bar \Lambda^k , && \Tilde{B}(1) = I,
        \\
        & \Tilde{H}(N) = \Omega^N, && \Tilde{H}(1) = \Omega,
        \\
        & \Tilde{M}(N) = \sum_{k=0}^{N-1} \left(\Omega^k \right)' , && \Tilde{M}(1) = I_h,
        \\
        & \Tilde{Q}(N) = \sum_{k=0}^{N-1} \left( \Omega^k \right)' \Tilde Q_c \left( \Omega^k \right), && \Tilde{Q}(1) = \Tilde Q_c,
        \\
        & \Tilde R(N) = \sum_{k=0}^{N-1} \left( \Lambda^k \right) \bar R_{ww,c} \left( \Lambda^k \right)', && \Tilde{R}(1) = \bar R_{ww,c},
    \end{alignat}
\end{subequations}
that can be used for computing 
\begin{subequations}
\label{eq:Stepdoubling_numericalExpressions}
    \begin{alignat}{5}
        & A(T_s) = \Tilde A(N)(1:n_x,1:n_x),  
        \; H(T_s) = \tilde H(N),
        \\
        & B_o(T_s) = \Theta_o \Tilde B_o(N) \bar B_{oc},
        \quad \; M(T_s) = E_{2}'\Tilde M(N)' \Tilde M_c,
        \\
        & Q = E_2' \Tilde Q(N) E_2, \qquad \quad \; R_{ww} = h \sum_{i=1}^s b_i \Lambda_i \Tilde R(N) \Lambda_i',
    \end{alignat}
\end{subequations}
where 
\begin{subequations}
\label{eq:StepdoublingCoefficients}
\begin{equation}
    \bar{\Lambda} = \text{diag}(\Lambda, \Lambda_v), \,
    \Theta_o = \begin{bmatrix}
        \Theta_1 & \Theta_2
    \end{bmatrix}, \,
    \bar{B}_{oc}= [\bar B_{1c}; \bar{B}_{2c}],
\end{equation}
\begin{alignat}{5}
    \Tilde M_c = h \sum_{i=1}^s b_i \Omega_i' E_1' \bar M_c, \:
    \Tilde Q_c = h \sum_{i=1}^s b_i \Omega_i' E_1' \bar Q_c E_1 \Omega_i.
\end{alignat}    
\end{subequations}
\cite{ScalingSquaringRevisited2005Nigham,NewStepDoubling2010Nigham,ExponentialIntegrators2011Higham} described the squaring and scaling algorithm for solving the matrix exponential problem. We can apply the idea of repeated squaring for computing matrices introduced in~\eqref{eq:stepDoublingMatices}, and it leads to the step-doubling method.  
\begin{table}[tb]
\begin{center}
\caption{The step-doubling functions} %
\begin{tabular}{ccc}
 & Numerical expression & Step-doubling function  \\ \hline
$\Tilde A(N)$ & $\Lambda^N$ & $\Tilde{A}(\frac{N}{2})\Tilde A (\frac{N}{2})$ \\
$\Tilde B(N)$ & $ \sum_{i=0}^{N-1}\bar \Lambda^i$ & $\Tilde B(\frac{N}{2})\left( I + \bar A(\frac{N}{2}) \right)$ \\ 
$\Tilde H(N)$ & $ \Omega^N$ & $\Tilde{H}(\frac{N}{2})\Tilde H (\frac{N}{2})$ \\
$\Tilde M(N)$ & $\sum_{i=0}^{N-1} \Omega^i$ & $\Tilde{M}(\frac{N}{2}) \left( I + \Tilde H(\frac{N}{2}) \right)$ \\ 
$\Tilde Q(N)$ & $\sum_{i=0}^{N-1} \left( \Omega^i \right)' \Tilde Q_c \left( \Omega^i\right) $ & $\Tilde Q(\frac{N}{2}) + \Tilde H(\frac{N}{2})' \Tilde Q(\frac{N}{2}) \Tilde H(\frac{N}{2})$ \\ 
$\Tilde R(N)$ & $\sum_{i=0}^{N-1} \left( \Lambda^i \right) \bar R_{ww,c} \left( \Lambda^i\right)' $ & $\Tilde R(\frac{N}{2}) + \Tilde A(\frac{N}{2}) \Tilde R(\frac{N}{2}) \Tilde A(\frac{N}{2})'$ \\ \hline
\end{tabular} %
\label{tb:stepDoublingExpressions} 
\end{center}
\end{table}
\begin{algorithm}[tb]
\caption{Step-doubling meth. for LQ Discretization}
\label{algo:Stepdoubling-LQDiscretization}
\begin{flushleft}
    \textbf{Input:} $(A_c, B_c, G_c, C_c, D_c, Q_c, T_s, j)$ \\
    \textbf{Output:} $(A,B_o,Q,M,R_{ww})$ 
\end{flushleft}
\begin{algorithmic}
\State Compute the integration step $N = 2^j$ 
\State Compute the step size $h = \frac{T_s}{N}$
\State Use~\eqref{eq:ODEmethods-StageVariableCoefficients} and~\eqref{eq:HkandGammak} to compute ($\Lambda_i$, $\Lambda_{v,i}$, $\Omega_i$, $\Theta_{1,i}$, $\Theta_{2,i}$)
\State Use~\eqref{eq:ODEmethods-ConstantCoefficients} and~\eqref{eq:HkandGammak} to compute ($\Lambda$, $\Lambda_{v}$, $\Omega$, $\Theta_{1}$, $\Theta_{2}$) 
\State Use~\eqref{eq:StepdoublingCoefficients} to compute ($\bar{\Lambda}$, $\Theta_o$, $\bar{B}_{oc}$, $\tilde M_c$, $\tilde Q_c$)
\State Set initial states ($i=1$, $\Tilde{A}(i) = \bar{\Lambda}$, $\Tilde{B}(i) = I$, \\ $\tilde H(i) = \Omega$, $\Tilde{Q}(i) = \Tilde{Q}_c$, $\Tilde{M}(i) = I$, $\Tilde{R}(i) = \bar{R}_{ww,c}$)
\While{$i \leq j$} 
    \State Use step-doubling functions~\eqref{eq:StepdoublingEquations} to update \\ $\qquad$ ($\Tilde{A}(i)$, $\Tilde A_v(i)$, $\Tilde{B}(i)$, $\tilde H(i)$, $\Tilde{Q}(i)$, $\Tilde{M}(i)$, $\Tilde{R}(i)$)
    \State Set $i \leftarrow i + 1$
\EndWhile
\State Use~\eqref{eq:Stepdoubling_numericalExpressions} to compute ($A$, $B_o$, $Q$, $M$, $R_{ww}$)
\end{algorithmic}
\end{algorithm}

Let $f(n)$ for $f \in \left[\Tilde A, \Tilde B_o, \Tilde H, \Tilde M, \Tilde Q, \Tilde R \right]$ represents functions described in~\eqref{eq:stepDoublingMatices}, and the integration step $N=2^j$ for $j \in \mathbb{Z}^+$. The step-doubling expression of $f(n)$ can be written as  
\begin{subequations}
\label{eq:StepdoublingEquations}
 \begin{equation}
    f(1) \rightarrow f(2) \rightarrow f(4) \rightarrow \cdots \rightarrow f(\frac{N}{2}) \rightarrow f(N),
\end{equation}
and
\begin{equation}
    f(n) = F(f(\frac{n}{2})), \qquad n \in \left[2,4,\ldots, \frac{N}{2}, N \right].
\end{equation}   
\end{subequations}
$F(x)$ is the step-doubling function for computing $f(n)$, and it takes the $\frac{n}{2}^{\text{th}}$ step's result to compute the double step's result $f(n)$. Consequently, we only take j steps to get the same results as the ODE method with N integration steps. Table~\ref{tb:stepDoublingExpressions} describes step-doubling functions, and Algorithm~\ref{algo:Stepdoubling-LQDiscretization} describes the step-doubling method for solving proposed differential equations systems.

\section{Numerical experiments}
\label{sec:NumericalExperiments}
This section presents numerical experiments for comparing proposed numerical discretization methods and testing the CT-MPC. 

The numerical experiment considers the cement mill system introduced by~\cite{Olesen2013ATuningProcedure} and~\cite{PRASATH2010ApplicationofSoftConstrainedMPC}, and it can be described as 
\begin{equation}
\label{eq:inputOutputmodel}
    \bs Y(s) = G_u(s) U(s) + G_d(s)(D(s) + \bs W(s)) + \bs V(s),
\end{equation}
with the transfer functions  
\begin{equation}
    G_u(s) = \begin{bmatrix}
        \frac{12.8e^{-s}}{16.7s+1} & \frac{-18.9e^{-3s}}{21.0s+1} \\ 
        \frac{6.6e^{-7s} }{10.9s+1}& \frac{-19.4e^{-3s} }{14.4s+1}
    \end{bmatrix},
    G_d(s) = \begin{bmatrix}
        \frac{-1.0e^{-3s}}{(32s+1)(21s+1)} \\ \frac{60}{(30s+1)(20s+1)}
    \end{bmatrix},
\label{eq:transferFunctionModel}
\end{equation}
where the inputs $u_1$ = feed flow rate [TPH] and $u_2$ = separator speed [\%] and the outputs $z_1$ = elevator load [kW] and $z_2$ = fineness [cm$^2$/g]. The disturbance $D$ is the clinker hardness [HGI]. $\bs W$ and $\bs V$ are the process and measurement noise.

The cement mill system~\eqref{eq:inputOutputmodel} is converted into a discrete-time state space model with $T_s = 2$ [min],
\begin{subequations}
    \begin{alignat}{3}
        \bs x_{k+1} & = A \bs x_k + B u_k + E d_k + G \bs w_k, \\
        \bs y_k &= C \bs x_k + D u_k + \bs v_k.
    \end{alignat}
\end{subequations}
We use the discrete-time state space model as the simulator in the numerical test. The covariance matrices for the process noise $\bs w_k$ and the measurement noise $\bs v_k$ are selected as $R_{ww}=1.0$ and $R_{vv}=\text{diag}(0.1,50)$. The simulation time is $T_{sim}=12$ [h] and the system steady states are $u_s = \left[128; 60\right]$ and $z_s = \left[25; 3100\right]$. The disturbance $d_k = 20$ for $t \in \set{3, 9}$ [h] and $d_k = 0$ for the rest time. 

\subsection{Discretization of CT-MPC}
The control model of the CT-MPC is $ \bs Z(s) = G(s) U(s) + H(s) \bs W(s)$. Here the deterministic model $G(s)$ is identical to $G_u(s)$ and $H(s)$ is the stochastic part of the control model and 
\begin{equation}
     H(s) = \text{diag} \left( \left[ \frac{1}{s} \frac{1}{10s+1}; \frac{1}{s} \frac{1}{10s+1} \right] \right).
\end{equation}
The transfer function models are converted into a continuous-time state space model. We use the continuous-time LQ-OCP introduced in~\eqref{eq:Cont-Stc-LQ-OCP} as the objective function of the CT-MPC. The weight matrix $Q_c$ = diag$\left(1.0, 1.0\right)$ and the control and prediction horizons are $N$ = $200$ [min].  
\begin{table}[tb]
\centering
    \caption{CPU time and error of the scenario using classic RK4 with $N=2^{14}$}%
    \label{tab:NumericalExperiment-ErrorsandCPUTime}
    \begin{tabular}{ccccc}
    \hline
                & & Matrix Exp. & \multicolumn{1}{l}{ODE} & \multicolumn{1}{l}{Step-doubling} \\ \hline
    $e(A)$      & [-] & -           & $1.03\cdot 10^{-12}$          & $1.03 \cdot 10^{-12}$             \\
    $e(B_o)$      & [-] & -           & $2.31 \cdot 10^{-12}$          & $2.31 \cdot 10^{-12}$             \\
    $e(R_{ww})$ & [-] & -           & $3.43 \cdot 10^{-12}$          & $3.43 \cdot 10^{-12}$             \\
    $e(M)$      & [-] & -           & $4.76 \cdot 10^{-7}$          & $4.76 \cdot 10^{-7}$             \\
    $e(Q)$      & [-] & -           & $5.51 \cdot 10^{-7}$          & $5.51 \cdot 10^{-7}$             \\
    CPU Time    & [s] & 0.29     & 2.94 & 0.03      \\ \hline
    \end{tabular}
\end{table}

We discretize the continuous-time LQ-OCP using the proposed three numerical methods. Table~\ref{tab:NumericalExperiment-ErrorsandCPUTime} illustrates the CPU time and error of the fixed-time-step ODE and step-doubling methods using the classic RK4 with $N=2^{14}$. We consider the results from the matrix exponential method as the true solution and the error is computed as $e(i)= \norm{i(T_s)-i(N)}_{\infty}$ for $i \in [A,B_o,R_{ww},M,Q]$. From Table~\ref{tab:NumericalExperiment-ErrorsandCPUTime}, we notice that the step-doubling method is the fastest among all three methods, spending only 3 [ms] while keeping the same accuracy as the fixed-time-step ODE method. 

\subsection{Closed-loop simulation}
Consequently, we obtain the discrete-time equivalent~\eqref{eq:analyticMeanOfStochasticCosts01}. Define the input sequence $u$ = $\left[u_0;u_1;\ldots,u_{N-1}\right]$ with $u_k=I_ku$, and we have 
\begin{equation}
    b_k= A^k x_0, \; \; \Gamma_k = \sum_{i=0}^k A^{k-1-i} B I_i, \; \; x_k = b_k + \Gamma_k u.
\end{equation}
Replacing $x_k$ and $u_k$ with the above expressions, we then get the following quadratic program (QP)
\begin{subequations}
\label{eq:convexQP}
\begin{alignat}{5}
    & \min_{ \{ u_k \}^{N-1}_{k=0} } \quad && \phi = \frac{1}{2} u'H u + g' u \\
    & s.t. && u_{min} \leq u_k \leq u_{max}, \quad && k \in \mathcal{N}, \\
    & && \Delta u_{min} \leq \Delta u_{k} \leq \Delta u_{max}, \quad && k \in \mathcal{N},
\end{alignat}
\end{subequations}
with the quadratic and linear terms matrices 
\begin{align}
    & H = \sum_{k\in \mathcal{N}}^{N-1} \begin{bmatrix}
        \Gamma_k \\ I_k
    \end{bmatrix}' Q \begin{bmatrix}
        \Gamma_k \\ I_k
    \end{bmatrix},
    g = \sum_{k=0}^{N-1} \begin{bmatrix}
        \Gamma_k \\ I_k
    \end{bmatrix}' \left( Q \begin{bmatrix}
        b_k \\ 0
    \end{bmatrix} + q_k \right),
\end{align}
where $q_k = M \bar z_k$.  $u_{min}=-20$, $u_{max}=20$ and $\Delta u_{min}= -2.0$, $\Delta u_{max}=2.0$ are input box and input rate-of-movement (ROM) constraints.
\begin{figure}
\begin{center}
\includegraphics[width=8.8cm]{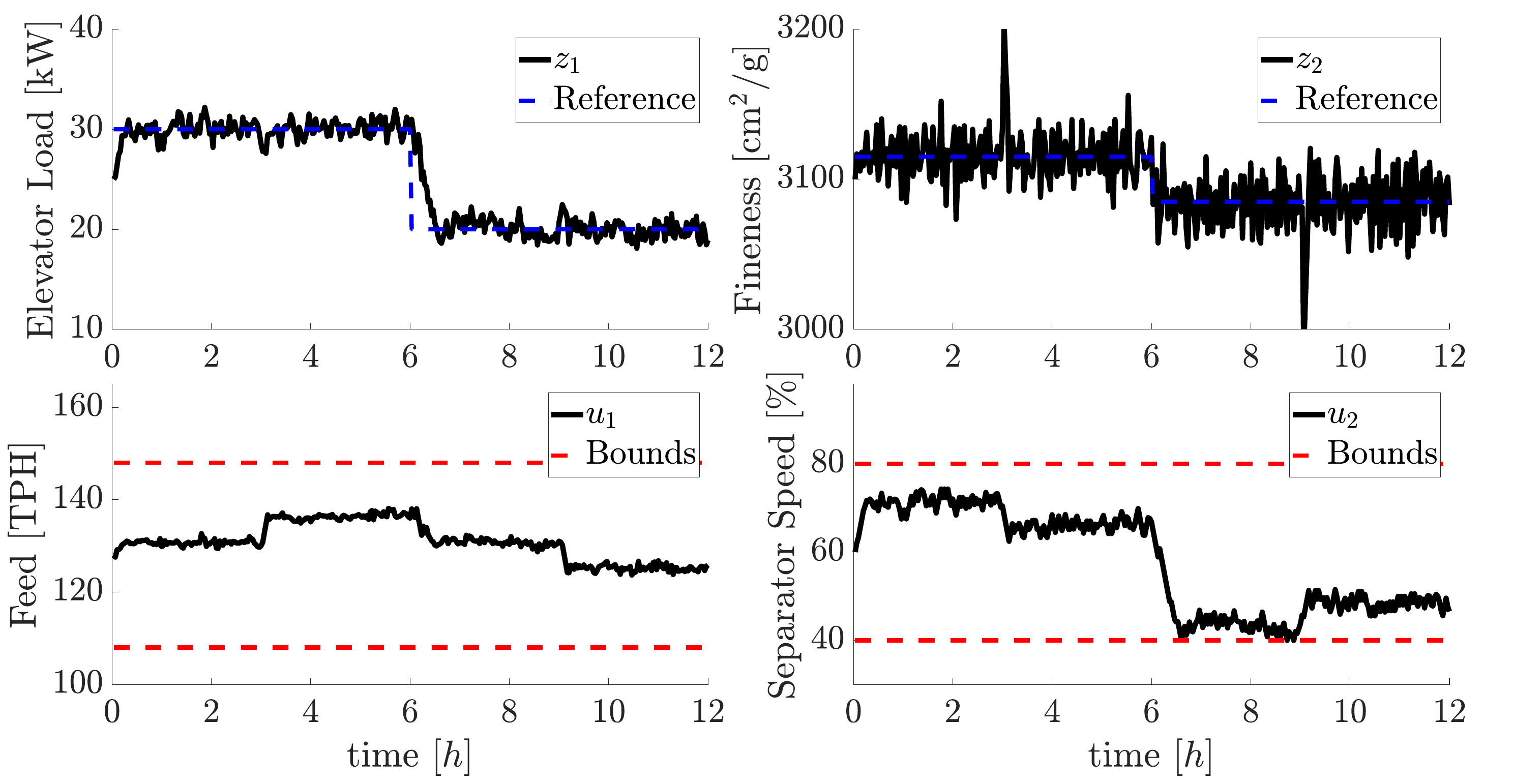}    
\caption{The closed-loop simulation of a simulated cement mill controlled by a LQ discretization-based MPC.} 
\label{fig:ClosedLoopSimulationCementMill}
\end{center}
\end{figure}

We implement the CT-MPC along with a linear Kalman filter on the simulated cement mill system. The covariance matrix of the measurement noise is $R_{vv} = \left[0.1; 50 \right]$ and the process noise covariance $R_{ww}$ is obtained from the differential equation $R_{ww}(T_s)$. The cross-covariance matrix is assumed to be $S=0$. Fig.~\ref{fig:ClosedLoopSimulationCementMill} illustrates the closed-loop simulation result of the cement mill system. Initially, the CT-MPC takes a while to bring two outputs to the reference (indicated by the blue lines). Then, there are overshoots on the outputs at $t = 3$ [h] caused by the disturbance $d_k$. The controller captures the unknown disturbance and rejects it after a few iterations. The references have a step change at $t = 6$ [h], and the system outputs are controlled to follow the new reference points. We withdraw the disturbance at $t = 9$ [h]. Thus, the overshoots appear again, and the controller successfully handles it.

\section{Conclusions}
\label{sec:Conclusions}
This paper introduced the discretization of continuous-time LQ-OCPs with time delays. We expressed the discrete weight matrices as the systems of differential equations, leading to the discrete equivalent of the continuous-time LQ-OCPs with time delays. Three numerical methods are described for solving proposed differential equation systems. We tested the CT-MPC with proposed numerical methods in the numerical experiment on a simulated cement mill system. The step-doubling is the fastest among all three methods and keeps the same accuracy level as the fixed-time-step ODE method. The closed-loop simulation results indicate that the proposed CT-MPCs can stabilize and control the simulated cement mill system.




\bibliography{reference}

\end{document}